\documentclass[aps,prl,twocolumn,superscriptaddress]{revtex4-1}
\usepackage{breakurl}
\usepackage{xcolor}
\usepackage{sidecap}
\usepackage{amssymb}
\usepackage{hhline}
\usepackage{multirow}
\sidecaptionvpos{figure}{t}
\usepackage{amsmath}
\usepackage{graphicx}
\usepackage{esint}
\usepackage{epstopdf}
\usepackage{rotating}
\graphicspath{{pict/}{./}}
\usepackage{bm}%
\usepackage{microtype,bm,bbm,graphicx,booktabs,times}

\newcounter{Fig}

\begin{document}

\title{Evolution and global charge conservation for polarization singularities emerging from nonhermitian degeneracies}
\author{Weijin Chen}
\affiliation{School of Optical and Electronic Information, Huazhong University of Science and Technology, Wuhan, Hubei 430074, P. R. China}
\author{Qingdong Yang}
\affiliation{School of Optical and Electronic Information, Huazhong University of Science and Technology, Wuhan, Hubei 430074, P. R. China}
\author{Yuntian Chen}
\email{yuntian@hust.edu.cn}
\affiliation{School of Optical and Electronic Information, Huazhong University of Science and Technology, Wuhan, Hubei 430074, P. R. China}
\affiliation{Wuhan National Laboratory for Optoelectronics, Huazhong University of Science and Technology, Wuhan, Hubei 430074, P. R. China}
\author{Wei Liu}
\email{wei.liu.pku@gmail.com}
\affiliation{College for Advanced Interdisciplinary Studies, National University of Defense
Technology, Changsha, Hunan 410073, P. R. China}
\begin{abstract}
Core concepts in singular optics, especially the polarization singularity, have rapidly penetrated the surging fields of topological and nonhermitian photonics. For open photonic structures with degeneracies in particular, the polarization singularity would inevitably encounter another sweeping concept of Berry phase. Several investigations have discussed, in an inexplicit way, the connections between both concepts, hinting at that nonzero topological charges for far-field polarizations on a loop is inextricably linked to its nontrivial Berry phase when degeneracies are enclosed. In this work, we reexamine the seminal photonic crystal slab that supports the fundamental two-level nonhermitian degeneracies.  Regardless of the invariance of nontrivial Berry phase for different loops enclosing both exceptional points, we demonstrate that the associated polarization fields exhibit topologically inequivalent patterns that are characterized by variant topological charges, including even the trivial scenario of zero charge. It is further revealed that for both bands, the seemingly complex evolutions of polarizations are bounded by the global charge conservation, with extra points of circular polarizations  playing indispensable roles. This indicates that tough not directly associated with any local charges, the invariant Berry phase is directly linked to the globally conserved charge, the physical principles underlying which have all been further clarified by a modified Berry-Dennis model. Our work can potentially trigger an avalanche of studies to explore subtle interplays between Berry phase and all sorts of optical singularities, shedding new light on subjects beyond photonics that are related to both Berry phase and singularities.
\end{abstract}
\maketitle

Pioneered by Pancharatnam, Berry, Nye and  others~\cite{PANCHARATNAM_1955_ProcIndianAcadSci_propagation,PANCHARATNAM_1956_ProcIndianAcadSci_Generalized,BERRY_1984_Proc.R.Soc.A_Quantal,berry_adiabatic_1987,BERRY_2010_Nat.Phys._Geometric,BERRY_1976_Adv.Phys._Waves,NYE_1974_Proc.R.Soc.Lond.Math.Phys.Sci._Dislocations,
NYE_1983_ProcRSocA_Polarization,NYE_1983_Proc.R.Soc.A_Lines,BERRY_2001_SecondInt.Conf.Singul.Opt.Opt.VorticesFundam.Appl._Geometry},  Berry phase and singularities have become embedded languages all across different branches of photonics. Optical Berry phase is largely manifested through either polarization evolving Pancharatnam-Berry phase or the spin-redirection Bortolotti-Rytov-Vladimirskii-Berry phase~\cite{PANCHARATNAM_1956_ProcIndianAcadSci_Generalized,berry_adiabatic_1987,BORTOLOTTI_1926_RendRAccNazLinc_Memories,RYTOV_1938_DoklAkadNaukSSSR_Transition,VLADIMIRSKII_1941_DoklAkadNaukSSSR_rotation,BERRY_2010_Nat.Phys._Geometric,BLIOKH_2019_Rep.Prog.Phys._Geometric,COHEN_2019_NatRevPhys_Geometric}; while optical singularities are widely observed as singularities of intensity (caustics)~\cite{BERRY_1976_Adv.Phys._Waves}, phase (vortices)~\cite{NYE_1974_Proc.R.Soc.Lond.Math.Phys.Sci._Dislocations} or polarization~\cite{NYE_1983_ProcRSocA_Polarization,NYE_1983_Proc.R.Soc.A_Lines,BERRY_2001_SecondInt.Conf.Singul.Opt.Opt.VorticesFundam.Appl._Geometry}. As singularities for complex vectorial waves, polarization singularities are skeletons of electromagnetic waves and are vitally important for understanding various interference effects underlying many applications~\cite{DENNIS_2009_ProgressinOptics_Chapter,GBUR_2016__Singular}.

There is a superficial similarity between the aforementioned two concepts: both the topological charge  of polarization field (Hopf index of line field~\cite{HOPF_2003__Differential}) and Berry phase are defined on a closed circuit. In spite of this, it is quite unfortunate that almost no definite connections have been established between them in optics. This is fully understandable: Berry phase is defined on the Pancharatnam connection (parallel transport) that decides the phase contrast between neighbouring states on the loop~\cite{BERRY_1984_Proc.R.Soc.A_Quantal,berry_adiabatic_1987};  while the polarization charge reflects accumulated orientation rotations of polarization ellipses, which has no direct relevance to overall phase of each state. This explains why in pioneering works where both concepts were present~\cite{BERRY_CURRENTSCIENCE-BANGALORE-_pancharatnam_1994,BERRY_2000_Nature_Making,BERRY_2003_Proc.R.Soc.Lond.A_optical,BERRY_2004_CzechoslovakJournalofPhysics_Physicsa,BERRY_2007_ProgressinOptics_Chapter}, their interplay were rarely elaborated.

Spurred by studies into bound states in the continuum, polarization singularities have gained enormous renewed interest in open periodic photonic structures, manifested in different morphologies with both generic and higher-order half-integer charges~\cite{HSU_Nat.Rev.Mater._bound_2016,HSU_Nature_observation_2013-1,ZHEN_2014_Phys.Rev.Lett._Topological,YANG_2014_Phys.Rev.Lett._Analytical,GUO_Phys.Rev.Lett._topologically_2017,
KODIGALA_Nature_lasing_2017,BULGAKOV_2017_Phys.Rev.Lett._Topological,DOELEMAN_2018_Nat.Photonics_Experimentala,ZHOU_2018_Science_Observationa,ZHANG_2018_Phys.Rev.Lett._Observation,KOSHELEV_2018_Phys.Rev.Lett._Asymmetrica,CHEN_2019__Singularities,CHEN_2019_Phys.Rev.B_Observing,LIU_2019_Phys.Rev.Lett._Circularly,JIN_2019_Nature_Topologicallya,
GUO_2020_Phys.Rev.Lett._Meron,YIN_2020_Nature_Observationa,YE_2020_Phys.Rev.Lett._Singular,CHEN_2019_ArXiv190409910Math-PhPhysicsphysics_Linea,LIU_2019_Phys.Rev.Lett._High,HUANG_2020_Science_Ultrafasta,WANG_2019_ArXiv190912618Phys._Generating}.
Simultaneously, the significance of Berry phase has been further reinforced in surging fields of topological and nonhermitian photonics~\cite{Lu2014_topological,OZAWA_2018_ArXiv180204173,PANCHARATNAM_1955_ProcIndianAcadSci_propagation,BERRY_CURRENTSCIENCE-BANGALORE-_pancharatnam_1994,BERRY_2004_CzechoslovakJournalofPhysics_Physicsa,FENG_2017_Nat.Photonics_NonHermitiana,EL-GANAINY_2018_Nat.Phys._NonHermitian,MIRI_2019_Science_Exceptionala}.  In periodic structures involving band degeneracies, Berry phase and polarization singularity would inevitably meet, which sparks the influential work on nonhermitian degeneracy~\cite{ZHOU_2018_Science_Observationa} and several other following studies~\cite{CHEN_2019_Phys.Rev.B_Observing,GUO_2020_Phys.Rev.Lett._Meron,YE_2020_Phys.Rev.Lett._Singular} discussing both concepts simultaneously. Though not claimed explicitly, those works hint that nontrivial Berry phase produces nonzero polarization charge.

Aiming to bridge Berry phase and polarization singularity, we reexamine the seminal photonic crystal slab (PCS) that supports elementary two-level nonhermitian degeneracies. Despite the invariance of nontrivial Berry phase, the corresponding polarization fields on different isofrequency contours enclosing both exceptional points (EPs) exhibit diverse patterns characterized by different polarization charges, including the trivial zero charge. It is further revealed such complexity of field evolutions is regulated by global charge conservation for both bands, with extra points of circular polarizations (\textbf{C}-points) playing pivotal roles. This reveals the explicit connection between globally conserved charge and the invariant Berry phase, underlying which the physical mechanisms have been further clarified by a modified Berry-Dennis model~\cite{BERRY_2003_Proc.R.Soc.Lond.A_optical}. Our study can spur further investigations in other subjects beyond photonics to explore conceptual  interconnectedness, where both the concepts of Berry phase and singularities are present.

\begin{figure}[tp]
\centerline{\includegraphics[width=9cm]{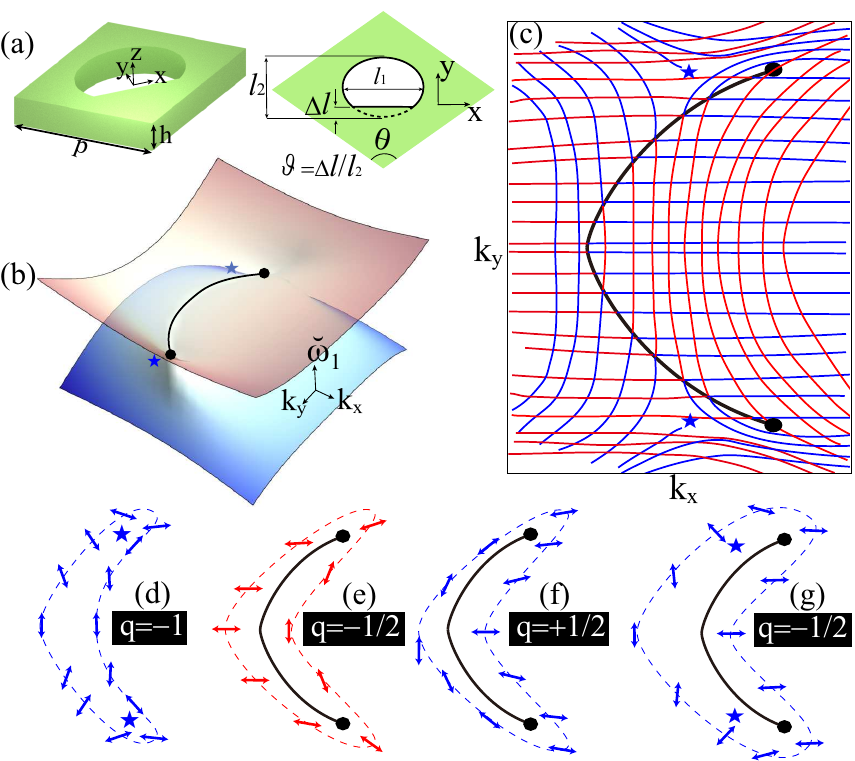}} \caption{\small  (a) Unit cell of the rhombic-lattice PCS: index $n=1.384$, $p=525$~nm, $h=220$~nm, $l_1=348$~nm, $l_2=257$~nm, $\theta=114.5^{\circ}$ and $\vartheta={\vartriangle}l/l_2$. (b) Dispersion bands ($\vartheta=0$) with two EPs $\breve{\omega}_1=0.961361$, $k_x=0.029525$, $k_y=\pm6.8\times10^{-4}$) and two \textbf{C}-points on the lower band $\breve{\omega}_1=0.961347$, $k_x=0.029517$, $k_y=\pm6.8\times10^{-4}$. The polarization fields on a loop enclosing two \textbf{C}-points  are shown in (d) with $q=-1$. (c) Polarization fields for both lower (blue) and upper (red) bands, and three isofrequency contours are selected ($\breve{\omega}_1=0.961368,0.961353,0.96133$), on which the polarization fields are summarized in (e)-(g), with $q=-1/2,~+1/2,~-1/2$, respectively. }
\label{fig1}
\end{figure}

For better comparisons, we revisit the rhombic-lattice PCS in Ref.~\cite{ZHOU_2018_Science_Observationa}: refractive index $n$, side length $p$, height $h$ and tilting angle $\theta$; semi-major (minor) diameters are $l_1$ ($l_2$); the whole structure is placed in air background of $n=1$ [Fig.~\ref{fig1}(a); parameter values shown in the figure caption].  We have further defined $\vartheta={\vartriangle}l/l_2$ to characterize the mirror ($k_y$-$k_z$ plane)-symmetry breaking  when air holes are partially filled. When $\vartheta=0$, dispersion bands (in terms of real parts of  complex eigenfrequencies $\breve{\omega}=\breve{\omega}_1+i\breve{\omega}_2$ for the Bloch eigenmodes calculated with COMSOL Multiphysics) are presented in Fig.~\ref{fig1}(b). Throughout this work, both frequency and wave vector are normalized: $\omega\rightarrow{\omega}p/2{\pi}c$ ($c$ is light speed); $\mathbf{k}\rightarrow\mathbf{k}p/2\pi$. Both branch cut (Fermi arc) and branch points (EPs) on the isofrequency plane (position information shown in figure captions, as is the case throughout this work) are observed [marked also respectively in Fig.~\ref{fig1}(c) by black curve and dots], confirming the existence of nonhermitian degeneracies.  On the lower band, we have identified two \textbf{C}-points (marked by stars; the corresponding eigenmodes are circularly polarized in the far field) on the isofrequency plane (position information shown in figure captions, as is the case throughout this work). Polarization fields (line fields in terms of the semi-major axis of the polarization ellipses) are projected on the Bloch vector $k_x$-$k_y$ plane [Fig.~\ref{fig1}(c)], with blue and red lines corresponding respectively to the eigenmodes on the lower and upper bands (fields exhibiting mirror symmetry as required by the structure symmetry). The representative eigenvalue-swapping feature is further confirmed in Fig.~\ref{fig1}(b), where the polarization fields are continuous across the Fermi arc for opposite bands only~\cite{BERRY_2003_Proc.R.Soc.Lond.A_optical}.

The coexistence of two \textbf{C}-points on the same band with equal charge $q=-1/2$ (generic polarization singularities) is protected by the mirror symmetry, decorated by typical star-like field patterns~\cite{BERRY_1977_J.Phys.A:Math.Gen._Umbilic}. 
On a contour that encloses two \textbf{C}-points (without enclosing EPs), the polarization fields are shown in Fig.~\ref{fig1}(d) with the expected charge $q=(-1/2)\times2=-1$.  Such a contour is not on an isofrequency plane and thus not quite feasible for direct experimental verifications.  We then proceed to isofrequency contours that are characterized by an invariant $\pi$ Berry phase~\cite{MAILYBAEV_2005_Phys.Rev.A_Geometric,LEYKAM_2017_Phys.Rev.Lett._Edge,SHEN_2018_Phys.Rev.Lett._Topological}. Since both  \textbf{C}-points locate on the lower bands and on the isofrequency plane: for the upper band, there is no \textbf{C}-point enclosed by the contour; for the lower band, the contour could enclose either zero or both \textbf{C}-points simultaneously.  Polarization fields on three such contours [one on the upper band (red dashed line) and two on the lower band (blue dashed lines)] are summarized in Figs.~\ref{fig1}(e)-(g), with $q=-1/2, +1/2, -1/2$, respectively. The charge contrast of $-1$ between the two contours on the lower band are obviously induced by \textbf{C}-points of total charge $q=-1$.

Though we have studied the same structure ($\vartheta=0$) as that in Ref.~\cite{ZHOU_2018_Science_Observationa}, our results presented in Fig.~\ref{fig1} are by no means mere reproductions, since the scenario of $q=+1/2$ we demonstrate is missing in Ref.~\cite{ZHOU_2018_Science_Observationa}, where the key roles of \textbf{C}-points are also overlooked. We emphasize that though not explicitly demonstrated, the case of $q=+1/2$ was actually not forbidden by the arguments presented in Ref.~\cite{ZHOU_2018_Science_Observationa}.  Based on mode swapping and mirror symmetry properties, it was proved there that the charge associated with the isofrequency contour has to be a half-integer, accommodating both $q=\pm1/2$.

\begin{figure}[tp]
\centerline{\includegraphics[width=9cm]{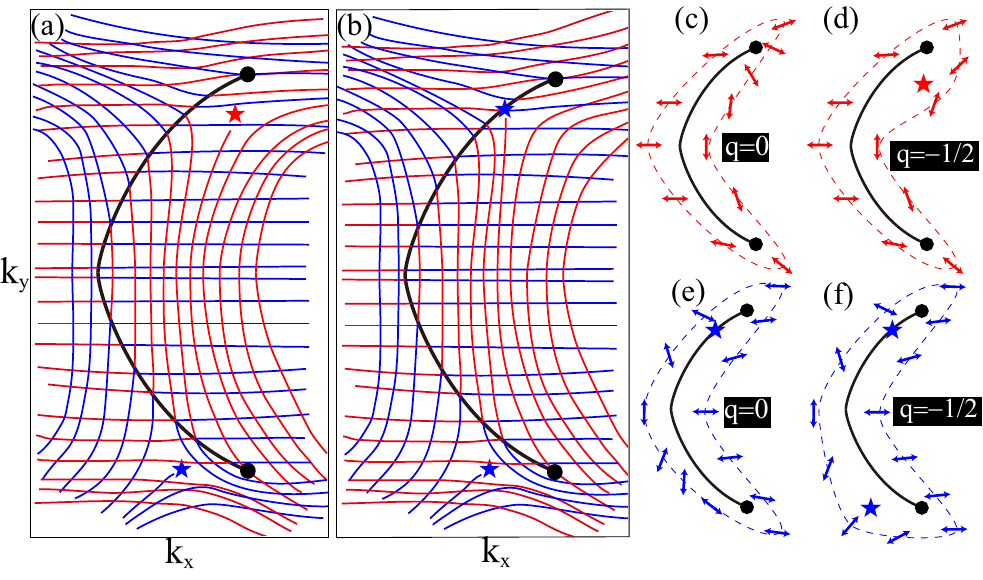}} \caption{\small (a) and (b) Polarization fields for two assymetric PCSs with  $\vartheta=0.01$ and $0.007$, respectively.  The positions for the EPs are $\breve{\omega}_1=0.9612505$, $k_x=0.029632$, $k_y=\pm6.4\times10^{-4}$ in (a) and $\breve{\omega}_1=0.961297$, $k_x=0.029592$, $k_y=\pm6.6\times10^{-4}$ in (b). The positons of the two \textbf{C}-points are: $\breve{\omega}_1=(0.961261,0.9612108)$, $k_x=(0.029638,0.029623)$, $k_y=(6.3\times10^{-4},-6.8\times10^{-4})$ in (a) and $\breve{\omega}_1=(0.961297,0.961270)$, $k_x=(0.029588,0.02958)$, $k_y=(6.48\times10^{-4},-6.8\times10^{-4})$ in (b).  In both (a) and (b), two isofrequency contours are chosen, on which the polarizations fields are shown in (c)-(f), with $q=0,-1/2,0,-1/2$ and $\breve{\omega}_1=0.9612$+ $(5.5,8.3,8.3,5)\times10^{-5}$, respectively.}
\label{fig2}  
\end{figure}

We then make a further step to investigate asymmetric structures ($\vartheta\neq0$).  The polarization fields  on the $k_x$-$k_y$ plane for two scenarios ($\vartheta=0.01,~0.007$) are summarized in Figs.~\ref{fig2}(a) and (b), neither exhibiting mirror symmetry anymore.  With symmetry broken, though one \textbf{C}-point on the lower band is relatively stable, the other can  move to the Fermi arc [Fig.~\ref{fig2}(b)] or across to the upper band [Figs.~\ref{fig2}(a)], with invariant $q=-1/2$ (see Table~\ref{table}).  When the two \textbf{C}-points locate on opposite bands [Fig.~\ref{fig2}(a)], we choose two contours on the upper band (the charge distribution on the lower band is similar): one encloses two EPs only and the other encloses also the \textbf{C}-point. The polarization fields on the contours are shown in Figs.~\ref{fig2}(c) and (d), with $q=0$ and $-1/2$, respectively. Despite this charge variance, we emphasize that for any isofrequency contour, the Berry phase is an invariant $\pi$, regardless of whether the symmetry is broken or not~\cite{MAILYBAEV_2005_Phys.Rev.A_Geometric,LEYKAM_2017_Phys.Rev.Lett._Edge,SHEN_2018_Phys.Rev.Lett._Topological}. Basically, Fig.~\ref{fig2}(c)  tells convincingly that a nontrivial Berry phase does not necessarily produce a nonzero polarization charge.

Except EPs, other points on the Fermi arc actually correspond to two sets of eigenmodes with equal $\breve{\omega}_1$  while different $\breve{\omega}_2$.  As a result, the \textbf{C}-point on the Fermi arc [Fig.~\ref{fig2}(b)] is not really shared by both bands (only EPs are shared), but still locate on the lower band, which can be confirmed by inspecting  $\breve{\omega}_2$. With the absence of \textbf{C}-points, the charge distribution on the upper band would be identical  to that in Fig.~\ref{fig1}(b): any isofrequency contour encloses two EPs only with $q=-1/2$. On the lower band, in contrast, an isofrequency contour can enclose either two EPs and inevitably a \textbf{C}-point on the Fermi arc, or two EPs and two \textbf{C}-points. Both scenarios are illustrated in Figs.~\ref{fig2}(e) and (f), with $q=0$ and $-1/2$, respectively. Figure~\ref{fig2}(e) reconfirms that Berry phase and partial polarization charge are not strictly interlocked.

Charge distributions for all three structures are summarized in Table~\ref{table}, with blank spaces corresponding to nonexistent scenarios.  Table~\ref{table} clearly indicates that for both the upper and lower bands, the global charge (when the contour is large enough to enclose both EPs and all \textbf{C}-points on the band) is invariant ($q=-1/2$), irrespective of how the \textbf{C}-points are distributed or whether the mirror symmetry is broken or not. In a word, there is a hidden order underlying the seemingly complex evolutions of  polarization fields and their charges:  the evolution is bounded by charge conservation. Considering the invariant $\pi$ Berry phase for any isofrequency contours, it becomes clear that the global polarization charge (rather than  partial ones when the contours covers part of the singularities of degeneracies or \textbf{C}-points) is inextricably linked to this invariant Berry phase. Such a subtle connection is also manifest for not only hermitian degeneracies~\cite{BERRY_2003_Proc.R.Soc.Lond.A_optical,BERRY_2007_ProgressinOptics_Chapter,YE_2020_Phys.Rev.Lett._Singular}, but also scenarios with the degeneracies removed by further perturbations~\cite{BERRY_2003_Proc.R.Soc.Lond.A_optical,GUO_2020_Phys.Rev.Lett._Meron}.

\begin{table}[htp]
\centering 
\begin{tabular}{|l|c|c|c|c|c|c|}
\hline \multirow{2}{*} {} & \multicolumn{2}{|c|} {$\vartheta=0$} & \multicolumn{2}{|c|} {$\vartheta=\mathrm{0.01}$} & \multicolumn{2}{|c|} {$\vartheta=\mathrm{0.007}$} \\
\cline { 2 - 7 } & \color{blue}{$\mathrm{L}$} & \color{red}{$\mathrm{U}$} & \color{blue}{$\mathrm{L}$} & \color{red}{$\mathrm{U}$} & \color{blue}{$\mathrm{L}$} & \color{red}{$\mathrm{U}$} \\
\hline ~~~~~~~~~~1C &  \color{blue}{$-1 / 2$} &  & \color{blue}{$-1 / 2$} & \color{red}{$-1 / 2$} &\color{blue}{$-1 / 2$} &  \\
\hline ~~~~Two EPs & \color{blue}{$+1 / 2$} & \color{red}{$-1 / 2$} & \color{blue}{0} & \color{red}{0} & & \color{red}{$-1/2$} \\
\hline Two EPs + 1C & & &  \color{blue}{$-1 / 2$} &  \color{red}{$-1 / 2$} &  \color{blue}{0} & \\
\hline Two EPs + 2Cs &  \color{blue}{$-1 / 2$} & & & &  \color{blue}{$-1 / 2$} & \\
\hline ~~~~~~~~Global &  \color{blue}{$-1 / 2$} & \color{red}{$-1 / 2$} & \color{blue}{$-1 / 2$} & \color{red}{$-1 / 2$} &\color{blue}{$-1 / 2$} & \color{red}{$-1 / 2$} \\
\hline
\end{tabular}\\
\caption{Charges for \textbf{C}-points and different isofrequency contours (\textbf{L}: lower band; \textbf{U}: Upper band). Blank spaces correspond to nonexistent scenarios.} 
\label{table} 
\end{table}

As the final step, we employ the local Berry-Dennis model proposed in Ref.~\cite{BERRY_2003_Proc.R.Soc.Lond.A_optical} to clarify the underlying mechanisms.  The corresponding Hamiltonian of this model in linear basis is:
\begin{equation}
\label{berry-dennis}
\mathcal{H}\left(k_{x}, k_{y}\right)=\left(k_{x}+i \gamma\right) \sigma_{z}+k_{y} \sigma_{x}+\kappa \sigma_{y},
\end{equation}
where $k_{x,y}$ are real; $\sigma_{x,y,z}$ are Pauli matrices; $\kappa$ and $\gamma$  are the planar chirality and radiation loss terms, respectively~\cite{Supplemental_Material_3}.   This Hamiltonian matrix is indeed a rather ordinary $2\times2$ nonhermition matrix, except that Berry and Dennis view its eigenvectors as Jones vectors~\cite{YARIV_2006__Photonics} for generally elliptically polarized light in linear basis,  thus establishing an effective connection between the Hamiltonian matrix and the electromagnetic polarization fields (see Supplemental Material (SM)~\cite{Supplemental_Material_3} for justifications of this connection and the incorporation of $\kappa$).   With this connection and the complex eigenvector denoted as $\textbf{x}=(x_1;x_2)$: when $\kappa=0$, EPs are chiral points with degenerate eigenvectors satisfying $x_1{\pm}ix_2=0$, overlapping with \textbf{C}-points; when $\kappa\neq0$, EPs are nonchiral and thus separated from \textbf{C}-points ~\cite{BERRY_2003_Proc.R.Soc.Lond.A_optical,HEISS_2001_Eur.Phys.J.D_chirality,HARNEY_2004_TheEuropeanPhysicalJournalD-AtomicMolecularandOpticalPhysics_Time,BERRY_2006_J.Phys.A:Math.Gen._Proximity}. Since for all the scenarios discussed above (see Figs.~\ref{fig1} and \ref{fig2}) the EPs do not overlap with \textbf{C}-points, the introduction of  chirality term $\kappa$ is inevitable, which is missing in Ref.~\cite{ZHOU_2018_Science_Observationa}.

\begin{figure}[tp]
\centerline{\includegraphics[width=9cm]{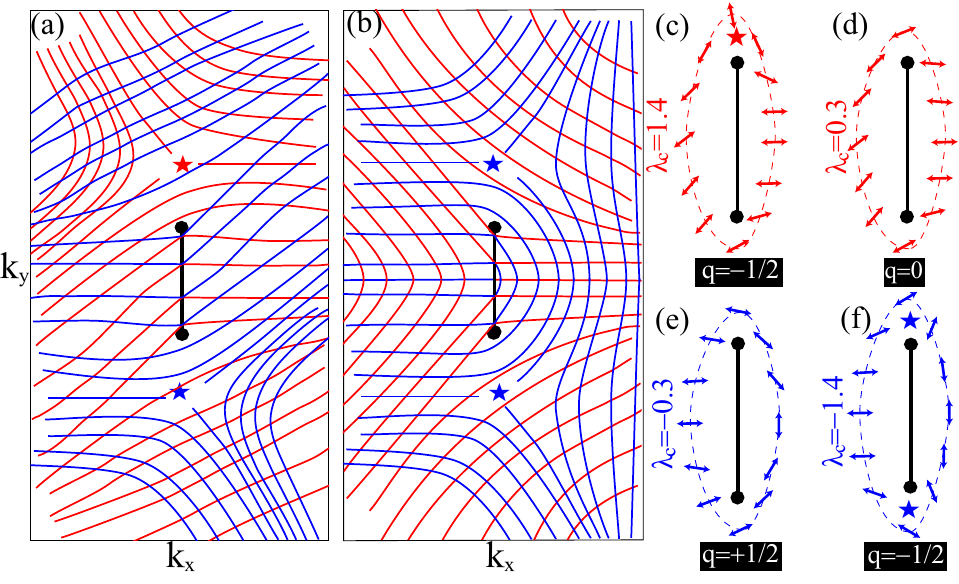}} \caption{\small (a) and (b) Polarization fields extrated from the model respectively in Eq.~(\ref{berry-dennis-circular2}) and Eq.~(\ref{berry-dennis-circular3}), with  $\gamma=1$ and $\kappa=0.8$. Two iso-eigenvalue contours are selected in (a) and (b), on which the polarizations fields are shown in (c)-(f), with $q=-1/2,0,0,-1/2$, respectively.}\label{fig3}
\end{figure}

For convenience of analysis, to directly locate \textbf{C}-points in particular, the Hamiltonian can be converted into a circular-basis form as~\cite{BERRY_2003_Proc.R.Soc.Lond.A_optical}:
\begin{equation}
\label{berry-dennis-circular}
\begin{aligned}
\mathcal{H}_c\left(k_{x}, k_{y}\right) &=\left(k_{x}+i \gamma\right) \sigma_{x}+k_{y} \sigma_{y}+\kappa \sigma_{z} \\
&=\left(\begin{array}{cc}
\kappa & k_{x}-\mathrm{i} k_{y}+\mathrm{i} \gamma \\
k_{x}+\mathrm{i} k_{y}+\mathrm{i} \gamma & -\kappa
\end{array}\right),
\end{aligned}\end{equation}
since such conversion would transform $\sigma_{x,y,z}$ in linear basis to $\sigma_{y,z,x}$ in circular basis~\cite{Supplemental_Material_3}. After this conversion, the chiral points now correspond to points of linear polarizations, while circular-basis eigenvectors of $x_1^cx_2^c=0$ correspond to \textbf{C}-points. Identical to the linear basis case, the EPs correspond to circular (noncircular) polarizations with the chirality term $\kappa=0$ ($\kappa\neq0$). The superiority of this circular-basis Hamiltonian resides in that the positions of \textbf{C}-points can then be directly identified by setting the off-diagonal terms of the matrix equal to zero:  $k_{x}-\mathrm{i} k_{y}+\mathrm{i} \gamma=0$  and $k_{x}+\mathrm{i} k_{y}+\mathrm{i} \gamma=0$. Their roots $k_x=0,~ k_y=\gamma$ and $k_x=0,~ k_y=-\gamma$ are the positions of \textbf{C}-points on the lower and upper bands, respectively~\cite{BERRY_2003_Proc.R.Soc.Lond.A_optical}. This model can explain the charge distributions shown in Fig.~\ref{fig2}(a) (also summarized in Table ~\ref{table} with $\vartheta=0.01$) with the two \textbf{C}-points located on opposite bands, except that in this model the topological charge of the \textbf{C}-point and the global charge for either band is $+1/2$ rather than $-1/2$ (see SM~\cite{Supplemental_Material_3}). To account for these discrepancies, we modify the Hamiltonian as:
\begin{equation}
\label{berry-dennis-circular2}
\mathcal{H}_c\left(k_{x}, k_{y}\right)=\left(k_{x}+i \gamma\right) \sigma_{x}-k_{y} \sigma_{y}+\kappa \sigma_{z},
\end{equation}
by adding a minus sign before the $\sigma_{y}$ term in Eq.~(\ref{berry-dennis-circular}).  This is similar to substituting the Hamiltonian of the K-valley for that of the $\rm{K}^{\prime}$-valley in graphene, which would induce a $2\pi$ jump of Berry phase from $\pi$ to $-\pi$~\cite{KANE_2005_Phys.Rev.Lett._Quantum,XIAO_2007_Phys.Rev.Lett._ValleyContrasting} (see SM~\cite{Supplemental_Material_3}). Though Berry phases of $\pi$ and $-\pi$ are effectively the same (phase is only definable modulo $2\pi$), the corresponding polarization fields and charge distributions are contrastingly different (see SM~\cite{Supplemental_Material_3} for the connections between Berry phase and polarization charges), with opposite sings for both the C-point charge and the global charge of both bands ($q=+1/2$ versus $q=-1/2$). The polarization fields extracted from this modified model ($\gamma=1$ and $\kappa=0.8$) are shown in Fig.~\ref{fig3}(a), which are topologically equivalent to those in Fig.~\ref{fig2}(a): for each \textbf{C}-point $q=-1/2$; for iso-eigenvalue ($\lambda_c$) contours that enclose both EPs, $q=0$ and $q=-1/2$ with and without the extra \textbf{C}-point surrounded, respectively [see Figs.~\ref{fig3}(c) and (d)]; the global charge is constant ($q=-1/2$) for both bands.

Since the model presented above is linear, there is only one solution  when either of the off-diagonal terms is setting to zero. This means that there is one and only one \textbf{C}-point on each band. As a result, this linear model would fail to account for what is observed in Fig.~\ref{fig1}(b), where there are two \textbf{C}-points on the same band. Actually the linear model in Eq.~(\ref{berry-dennis-circular2}) has broken the $k_x$-$k_z$ mirror symmetry of the polarization fields (symmetries of the Hamiltonian and constructed  fields are different, due to involvements of construction basis~\cite{Supplemental_Material_3}), as confirmed by the field patterns in Fig.~\ref{fig3}(a). To reflect the mirror symmetry of the structure and thus also the polarization fields, the linear model can be further modified as:
\begin{equation}
\label{berry-dennis-circular3}
\mathcal{H}_{c}\left(k_{x}, k_{y}\right)=\left(k_{x}+i \gamma\right) \sigma_{x}-k_{y} \sigma_{y}-k_{y}\sigma_{z},
\end{equation}
where the constant chirality term $\kappa$ in Eq.~(\ref{berry-dennis-circular2}) is replaced by the variable $-k_{y}$, which guarantees that the constructed polarization fields are symmetric with respect to the $k_y$-$k_z$ plane (see  SM~\cite{Supplemental_Material_3} for detailed arguments concerning field symmetry).  The symmetric fields [in contrast to asymmetric ones in Fig.~\ref{fig3}(a)] based on this model ($\gamma=1$ and $\kappa=0.8$) are shown in Fig.~\ref{fig3}(b), which is topologically equivalent to Fig.~\ref{fig1}(b): for each \textbf{C}-point $q=-1/2$; for iso-eigenvalue contours on the upper band $q=-1/2$; iso-eigenvalue contours on the lower band that enclose both EPs have $q=+1/2$ and $q=-1/2$, with and without the two \textbf{C}-points surrounded, respectively [see Figs.~\ref{fig3}(e) and (f)]; the global charge is an invariant $q=-1/2$ for both bands. This reconfirms the claim in Ref.~\cite{ZHOU_2018_Science_Observationa}: combined mirror symmetry and mode swapping produces half-integer charges. Here for simplicity we have confined to linear models only, aiming to explain topologically what has been observed in Figs.~\ref{fig1} and \ref{fig2}.  To obtain more than one \textbf{C}-points on the same band, besides introducing the variable chirality term as shown in Eq.~(\ref{berry-dennis-circular3}), we can also incorporate nonlinear terms into the Berry-Dennis model, which nevertheless would change the charge distribution both locally and globally (see SM~\cite{Supplemental_Material_3}).



In conclusion, we revisit PCSs supporting nonhermitian degeneracies and establish a subtle connection between invariant Berry phase and the conserved global charge. It is revealed that for any isofrequency contour enclosing both EPs, despite the  nontrivial $\pi$ Berry phase invariance, the topological charge is contrastingly variable, which could even be the trivial zero charge. Such seemingly complex evolutions of charge distributions are mediated by extra \textbf{C}-points, ensuring global charge conservation for both bands that is synonymous with the Berry phase invariance.

Our discussions are confined to fundamental two-level systems, which can be extended to more sophisticated systems with more complex EPs distributions~\cite{HEISS_2008_J.Phys.A:Math.Theor._Chirality,RYU_2012_Phys.Rev.A_Analysis,LEE_2012_Phys.Rev.A_Geometrica,HEISS_2015_J.Phys.A:Math.Theor._Resonance,LIN_2016_Phys.Rev.Lett._Enhanceda,HEISS_2016_J.Phys.A:Math.Theor._model,CHEN_2017_Nature_Exceptional,
HODAEI_2017_Nature_Enhanced,ZHANG_2018_Phys.Rev.X_Dynamically,PAP_2018_Phys.Rev.A_NonAbelian,DING_2018_Phys.Rev.Lett._Experimentala,ZHANG_2018_Phys.Rev.A_Hybrid,ZHONG_2018_Nat.Commun._Winding}. We emphasize that in this work, the Berry phase and polarization charge actually characterize different entities of  eigenvectors of Bloch modes and their projected far fields: Bloch modes are defined on the momentum-torus and can be folded into the irreducible Brillouin zone; while far fields are defined on the momentum-sphere, due to the involvement of out-of-plane wave vectors along which there is no periodicity. It is recently shown the Berry phase for electromagnetic fields themselves on a contour can be well defined~\cite{BLIOKH_2019_Rep.Prog.Phys._Geometric,BERRY_2019_J.Opt._Geometry}. We expect that blending all those  concepts (non-hermitian degeneracies, Berry phase of their matrix eigenvectors, Berry phase and polarization singularities of the corresponding electromagnetic waves) would render much more fertile platforms to incubate new fundamental investigations and practical applications, including the rare scenario of Berry phase (for electromagnetic fields) with slaving parameters (eigenvectors from which the electromagnetic fields are constructed) themselves also having Berry phase.

\emph{Acknowledgments}: We acknowledge the financial support from National Natural Science Foundation of China (Grant No. 11874026 and 11874426), and several other Researcher Schemes of National University of Defense Technology. W. L. is indebted to Sir Michael Berry and Prof. Tristan Needham for invaluable correspondences.

\begin{thebibliography}{10}
\newcommand{\enquote}[1]{``#1''}

\bibitem{PANCHARATNAM_1955_ProcIndianAcadSci_propagation}
S.~Pancharatnam, \enquote{The propagation of light in absorbing biaxial
  crystals,} Proc. Indian. Acad. Sci. \textbf{42}, 86--109 (1955).

\bibitem{PANCHARATNAM_1956_ProcIndianAcadSci_Generalized}
S.~Pancharatnam, \enquote{Generalized theory of interference, and its
  applications,} Proc. Indian. Acad. Sci. \textbf{44}, 247--262 (1956).

\bibitem{BERRY_1984_Proc.R.Soc.A_Quantal}
M.~V. Berry, \enquote{Quantal {{Phase Factors Accompanying Adiabatic
  Changes}},} Proc. R. Soc. A \textbf{392}, 45--57 (1984).

\bibitem{berry_adiabatic_1987}
M.~V. Berry, \enquote{The {{Adiabatic Phase}} and {{Pancharatnam Phase}} for
  {{Polarized}}-{{Light}},} J. Mod. Opt. \textbf{34}, 1401 (1987).

\bibitem{BERRY_2010_Nat.Phys._Geometric}
M.~Berry, \enquote{Geometric phase memories,} Nat. Phys. \textbf{6}, 148--150
  (2010).

\bibitem{BERRY_1976_Adv.Phys._Waves}
M.~V. Berry, \enquote{Waves and {{Thom}}'s theorem,} Adv. Phys. \textbf{25},
  1--26 (1976).

\bibitem{NYE_1974_Proc.R.Soc.Lond.Math.Phys.Sci._Dislocations}
J.~F. Nye and M.~V. Berry, \enquote{Dislocations in wave trains,} Proc. R. Soc.
  Lond. A \textbf{336}, 165--190 (1974).

\bibitem{NYE_1983_ProcRSocA_Polarization}
J.~F. Nye, \enquote{Polarization {{Effects}} in the {{Diffraction}} of
  {{Electromagnetic Waves}}: {{The Role}} of {{Disclinations}},} Proc. R. Soc.
  A \textbf{387}, 105--132 (1983).

\bibitem{NYE_1983_Proc.R.Soc.A_Lines}
J.~F. Nye, \enquote{Lines of circular polarization in electromagnetic wave
  fields,} Proc. R. Soc. A \textbf{389}, 279--290 (1983).

\bibitem{BERRY_2001_SecondInt.Conf.Singul.Opt.Opt.VorticesFundam.Appl._Geometry}
M.~V. Berry, \enquote{Geometry of phase and polarization singularities
  illustrated by edge diffraction and the tides,} in \enquote{Second
  {{International Conference}} on {{Singular Optics}} ({{Optical Vortices}}):
  {{Fundamentals}} and {{Applications}},} , vol. 4403 ({International Society
  for Optics and Photonics}, 2001), vol. 4403, pp. 1--12.

\bibitem{BORTOLOTTI_1926_RendRAccNazLinc_Memories}
E.~Bortolotti, \enquote{Memories and notes presented by fellows,} Rend. R. Acc.
  Naz. Linc. \textbf{4}, 552 (1926).

\bibitem{RYTOV_1938_DoklAkadNaukSSSR_Transition}
S.~M. Rytov, \enquote{Transition from wave to geometrical optics,} Dokl. Akad.
  Nauk. SSSR. \textbf{18}, 263 (1938).

\bibitem{VLADIMIRSKII_1941_DoklAkadNaukSSSR_rotation}
V.~V. Vladimirskii, \enquote{The rotation of polarization plane for curved
  light ray,} Dokl. Akad. Nauk. SSSR. \textbf{21}, 222 (1941).

\bibitem{BLIOKH_2019_Rep.Prog.Phys._Geometric}
K.~Y. Bliokh, M.~A. Alonso, and M.~R. Dennis, \enquote{Geometric phases in
  {{2D}} and {{3D}} polarized fields: Geometrical, dynamical, and topological
  aspects,} Rep. Prog. Phys. \textbf{82}, 122401 (2019).

\bibitem{COHEN_2019_NatRevPhys_Geometric}
E.~Cohen, H.~Larocque, F.~Bouchard, F.~Nejadsattari, Y.~Gefen, and E.~Karimi,
  \enquote{Geometric phase from {{Aharonov}}\textendash{{Bohm}} to
  {{Pancharatnam}}\textendash{{Berry}} and beyond,} Nat Rev Phys \textbf{1},
  437--449 (2019).

\bibitem{DENNIS_2009_ProgressinOptics_Chapter}
M.~R. Dennis, K.~O'Holleran, and M.~J. Padgett, \enquote{Chapter 5 {{Singular
  Optics}}: {{Optical Vortices}} and {{Polarization Singularities}},} in
  \enquote{Progress in {{Optics}},} , vol.~53, E.~Wolf, ed. ({Elsevier}, 2009),
  pp. 293--363.

\bibitem{GBUR_2016__Singular}
G.~J. Gbur, \emph{Singular {{Optics}}} ({CRC Press Inc}, Boca Raton, 2016).

\bibitem{HOPF_2003__Differential}
H.~Hopf, \emph{Differential {{Geometry}} in the {{Large}}: {{Seminar Lectures
  New York University}} 1946 and {{Stanford University}} 1956} ({Springer},
  2003).

\bibitem{BERRY_CURRENTSCIENCE-BANGALORE-_pancharatnam_1994}
M.~Berry, \enquote{Pancharatnam, virtuoso of the {{Poincar\'e}} sphere: An
  appreciation,} Curr. Sci. \textbf{67}, 220--220 (1994).

\bibitem{BERRY_2000_Nature_Making}
M.~Berry, \enquote{Making waves in physics,} Nature \textbf{403}, 21--21
  (2000).

\bibitem{BERRY_2003_Proc.R.Soc.Lond.A_optical}
M.~V. Berry and M.~R. Dennis, \enquote{The optical singularities of
  birefringent dichroic chiral crystals,} Proc. R. Soc. Lond. A \textbf{459},
  1261--1292 (2003).

\bibitem{BERRY_2004_CzechoslovakJournalofPhysics_Physicsa}
M.~Berry, \enquote{Physics of {{Nonhermitian Degeneracies}},} Czechoslovak
  Journal of Physics \textbf{54}, 1039--1047 (2004).

\bibitem{BERRY_2007_ProgressinOptics_Chapter}
M.~V. Berry and M.~R. Jeffrey, \enquote{Chapter 2 {{Conical}} diffraction:
  {{Hamilton}}'s diabolical point at the heart of crystal optics,} in
  \enquote{Progress in {{Optics}},} , vol.~50, E.~Wolf, ed. ({Elsevier}, 2007),
  pp. 13--50.

\bibitem{HSU_Nat.Rev.Mater._bound_2016}
C.~W. Hsu, B.~Zhen, A.~D. Stone, J.~D. Joannopoulos, and M.~Solja{\v c}i\'c,
  \enquote{Bound states in the continuum,} Nat. Rev. Mater. \textbf{1}, 16048
  (2016).

\bibitem{HSU_Nature_observation_2013-1}
C.~W. Hsu, B.~Zhen, J.~Lee, S.-L. Chua, S.~G. Johnson, J.~D. Joannopoulos, and
  M.~Solja{\v c}i\'c, \enquote{Observation of trapped light within the
  radiation continuum,} Nature \textbf{499}, 188--191 (2013).

\bibitem{ZHEN_2014_Phys.Rev.Lett._Topological}
B.~Zhen, C.~W. Hsu, L.~Lu, A.~D. Stone, and M.~Solja{\v c}i\'c,
  \enquote{Topological nature of optical bound states in the continuum,} Phys.
  Rev. Lett. \textbf{113}, 257401 (2014).

\bibitem{YANG_2014_Phys.Rev.Lett._Analytical}
Y.~Yang, C.~Peng, Y.~Liang, Z.~Li, and S.~Noda, \enquote{Analytical
  {{Perspective}} for {{Bound States}} in the {{Continuum}} in {{Photonic
  Crystal Slabs}},} Phys. Rev. Lett. \textbf{113}, 037401 (2014).

\bibitem{GUO_Phys.Rev.Lett._topologically_2017}
Y.~Guo, M.~Xiao, and S.~Fan, \enquote{Topologically {{Protected Complete
  Polarization Conversion}},} Phys. Rev. Lett. \textbf{119}, 167401 (2017).

\bibitem{KODIGALA_Nature_lasing_2017}
A.~Kodigala, T.~Lepetit, Q.~Gu, B.~Bahari, Y.~Fainman, and B.~Kant\'e,
  \enquote{Lasing action from photonic bound states in continuum,} Nature
  \textbf{541}, 196--199 (2017).

\bibitem{BULGAKOV_2017_Phys.Rev.Lett._Topological}
E.~N. Bulgakov and D.~N. Maksimov, \enquote{Topological {{Bound States}} in the
  {{Continuum}} in {{Arrays}} of {{Dielectric Spheres}},} Phys. Rev. Lett.
  \textbf{118}, 267401 (2017).

\bibitem{DOELEMAN_2018_Nat.Photonics_Experimentala}
H.~M. Doeleman, F.~Monticone, W.~den Hollander, A.~Al\`u, and A.~F. Koenderink,
  \enquote{Experimental observation of a polarization vortex at an optical
  bound state in the continuum,} Nat. Photonics \textbf{12}, 397 (2018).

\bibitem{ZHOU_2018_Science_Observationa}
H.~Zhou, C.~Peng, Y.~Yoon, C.~W. Hsu, K.~A. Nelson, L.~Fu, J.~D. Joannopoulos,
  M.~Solja{\v c}i{\'c}, and B.~Zhen, \enquote{Observation of bulk {{Fermi}} arc
  and polarization half charge from paired exceptional points,} Science
  \textbf{359}, 1009--1012 (2018).

\bibitem{ZHANG_2018_Phys.Rev.Lett._Observation}
Y.~Zhang, A.~Chen, W.~Liu, C.~W. Hsu, B.~Wang, F.~Guan, X.~Liu, L.~Shi, L.~Lu,
  and J.~Zi, \enquote{Observation of {{Polarization Vortices}} in {{Momentum
  Space}},} Phys. Rev. Lett. \textbf{120}, 186103 (2018).

\bibitem{KOSHELEV_2018_Phys.Rev.Lett._Asymmetrica}
K.~Koshelev, S.~Lepeshov, M.~Liu, A.~Bogdanov, and Y.~Kivshar,
  \enquote{Asymmetric {{Metasurfaces}} with {{High}}-{{Q}} {{Resonances
  Governed}} by {{Bound States}} in the {{Continuum}},} Phys. Rev. Lett.
  \textbf{121}, 193903 (2018).

\bibitem{CHEN_2019__Singularities}
W.~Chen, Y.~Chen, and W.~Liu, \enquote{Singularities and poincar\'e indices of
  electromagnetic multipoles,} Phys. Rev. Lett. \textbf{122}, 153907 (2019).

\bibitem{CHEN_2019_Phys.Rev.B_Observing}
A.~Chen, W.~Liu, Y.~Zhang, B.~Wang, X.~Liu, L.~Shi, L.~Lu, and J.~Zi,
  \enquote{Observing vortex polarization singularities at optical band
  degeneracies,} Phys. Rev. B \textbf{99}, 180101 (2019).

\bibitem{LIU_2019_Phys.Rev.Lett._Circularly}
W.~Liu, B.~Wang, Y.~Zhang, J.~Wang, M.~Zhao, F.~Guan, X.~Liu, L.~Shi, and
  J.~Zi, \enquote{Circularly {{Polarized States Spawning}} from {{Bound
  States}} in the {{Continuum}},} Phys. Rev. Lett. \textbf{123}, 116104 (2019).

\bibitem{JIN_2019_Nature_Topologicallya}
J.~Jin, X.~Yin, L.~Ni, M.~Solja{\v c}i{\'c}, B.~Zhen, and C.~Peng,
  \enquote{Topologically enabled ultrahigh- {{Q}} guided resonances robust to
  out-of-plane scattering,} Nature \textbf{574}, 501--504 (2019).

\bibitem{GUO_2020_Phys.Rev.Lett._Meron}
C.~Guo, M.~Xiao, Y.~Guo, L.~Yuan, and S.~Fan, \enquote{Meron {{Spin Textures}}
  in {{Momentum Space}},} Phys. Rev. Lett. \textbf{124}, 106103 (2020).

\bibitem{YIN_2020_Nature_Observationa}
X.~Yin, J.~Jin, M.~Solja{\v c}i{\'c}, C.~Peng, and B.~Zhen,
  \enquote{Observation of topologically enabled unidirectional guided
  resonances,} Nature \textbf{580}, 467--471 (2020).

\bibitem{YE_2020_Phys.Rev.Lett._Singular}
W.~Ye, Y.~Gao, and J.~Liu, \enquote{Singular {{Points}} of {{Polarizations}} in
  the {{Momentum Space}} of {{Photonic Crystal Slabs}},} Phys. Rev. Lett.
  \textbf{124}, 153904 (2020).

\bibitem{CHEN_2019_ArXiv190409910Math-PhPhysicsphysics_Linea}
W.~Chen, Y.~Chen, and W.~Liu, \enquote{Line {{Singularities}} and {{Hopf
  Indices}} of {{Electromagnetic Multipoles}},} Laser Photonics Rev., Doi:
  10.1002/lpor.202000049  (2020).

\bibitem{LIU_2019_Phys.Rev.Lett._High}
Z.~Liu, Y.~Xu, Y.~Lin, J.~Xiang, T.~Feng, Q.~Cao, J.~Li, S.~Lan, and J.~Liu,
  \enquote{High-{{Q}} {{Quasibound States}} in the {{Continuum}} for
  {{Nonlinear Metasurfaces}},} Phys. Rev. Lett. \textbf{123}, 253901 (2019).

\bibitem{HUANG_2020_Science_Ultrafasta}
C.~Huang, C.~Zhang, S.~Xiao, Y.~Wang, Y.~Fan, Y.~Liu, N.~Zhang, G.~Qu, H.~Ji,
  J.~Han, L.~Ge, Y.~Kivshar, and Q.~Song, \enquote{Ultrafast control of vortex
  microlasers,} Science \textbf{367}, 1018--1021 (2020).

\bibitem{WANG_2019_ArXiv190912618Phys._Generating}
B.~Wang, W.~Liu, M.~Zhao, J.~Wang, Y.~Zhang, A.~Chen, F.~Guan, X.~Liu, L.~Shi,
  and J.~Zi, \enquote{Generating optical vortex beams by momentum-space
  polarization vortices centered at bound states in the continuum,}
  arXiv190912618  (2019).

\bibitem{Lu2014_topological}
L.~Lu, J.~D. Joannopoulos, and M.~Soljacic, \enquote{Topological photonics,}
  Nat. Photonics \textbf{8}, 821 (2014).

\bibitem{OZAWA_2018_ArXiv180204173}
T.~Ozawa, H.~M. Price, A.~Amo, N.~Goldman, M.~Hafezi, L.~Lu, M.~C. Rechtsman,
  D.~Schuster, J.~Simon, O.~Zilberberg, and I.~Carusotto, \enquote{Topological
  photonics,} Rev. Mod. Phys. \textbf{91}, 015006 (2019).

\bibitem{FENG_2017_Nat.Photonics_NonHermitiana}
L.~Feng, R.~{El-Ganainy}, and L.~Ge, \enquote{Non-{{Hermitian}} photonics based
  on parity\textendash time symmetry,} Nat. Photonics \textbf{11}, 752--762
  (2017).

\bibitem{EL-GANAINY_2018_Nat.Phys._NonHermitian}
R.~{El-Ganainy}, K.~G. Makris, M.~Khajavikhan, Z.~H. Musslimani, S.~Rotter, and
  D.~N. Christodoulides, \enquote{Non-{{Hermitian}} physics and {{PT}}
  symmetry,} Nat. Phys. \textbf{14}, 11--19 (2018).

\bibitem{MIRI_2019_Science_Exceptionala}
M.-A. Miri and A.~Al{\`u}, \enquote{Exceptional points in optics and
  photonics,} Science \textbf{363}, eaar7709 (2019).

\bibitem{BERRY_1977_J.Phys.A:Math.Gen._Umbilic}
M.~V. Berry and J.~H. Hannay, \enquote{Umbilic points on {{Gaussian}} random
  surfaces,} J. Phys. A: Math. Gen. \textbf{10}, 1809--1821 (1977).

\bibitem{MAILYBAEV_2005_Phys.Rev.A_Geometric}
A.~A. Mailybaev, O.~N. Kirillov, and A.~P. Seyranian, \enquote{Geometric phase
  around exceptional points,} Phys. Rev. A \textbf{72}, 014104 (2005).

\bibitem{LEYKAM_2017_Phys.Rev.Lett._Edge}
D.~Leykam, K.~Y. Bliokh, C.~Huang, Y.~D. Chong, and F.~Nori, \enquote{Edge
  {{Modes}}, {{Degeneracies}}, and {{Topological Numbers}} in
  {{Non}}-{{Hermitian Systems}},} Phys. Rev. Lett. \textbf{118}, 040401 (2017).

\bibitem{SHEN_2018_Phys.Rev.Lett._Topological}
H.~Shen, B.~Zhen, and L.~Fu, \enquote{Topological {{Band Theory}} for
  {{Non}}-{{Hermitian Hamiltonians}},} Phys. Rev. Lett. \textbf{120}, 146402
  (2018).

\bibitem{Supplemental_Material_3}
Supplemental Material includes the following six sections:
  (\textbf{\uppercase\expandafter{\romannumeral1}}). The transformations of
  Pauli matrices with conversions from linear to circular basis;
  (\textbf{\uppercase\expandafter{\romannumeral2}}). Employment of Berry-Dennis
  model for the photonic crystal slab;
  (\textbf{\uppercase\expandafter{\romannumeral3}}). Berry phase around two EPs
  of the nonhermitian Hamiltonians in Eq.(2) and Eq.(3);
  (\textbf{\uppercase\expandafter{\romannumeral4}}). Global charge of the
  polarization fields constructed from nonhermitian Hamiltonians in Eq.(2) and
  Eq.(3); (\textbf{\uppercase\expandafter{\romannumeral5}}). The mirror
  asymmetry and symmetry for polarization fields constructed from nonhermitian
  Hamiltonians in Eqs.(2-3) and Eq.(4);
  (\textbf{\uppercase\expandafter{\romannumeral6}}). Introducing nonlinear
  terms into the linear model. Supplemental Material include the following
  Refs.~\cite{YARIV_2006__Photonics, BERRY_2003_Proc.R.Soc.Lond.A_optical,
  ZHOU_2018_Science_Observationa, PAPAKOSTAS_Phys.Rev.Lett._optical_2003,
  CHEN_2020_Phys.Rev.Research_Scatteringa,
  HARNEY_2004_TheEuropeanPhysicalJournalD-AtomicMolecularandOpticalPhysics_Time,
  BERRY_2006_J.Phys.A:Math.Gen._Proximity, MAILYBAEV_2005_Phys.Rev.A_Geometric,
  ZHOU_2016__Tailoring, LEYKAM_2017_Phys.Rev.Lett._Edge,
  SHEN_2018_Phys.Rev.Lett._Topological, BERRY_2007_ProgressinOptics_Chapter,
  CHEN_2019_ArXiv190409910Math-PhPhysicsphysics_Linea,
  BERRY_1977_J.Phys.A:Math.Gen._Umbilic}. 

\bibitem{YARIV_2006__Photonics}
A.~Yariv and P.~Yeh, \emph{Photonics: {{Optical Electronics}} in {{Modern
  Communications}}} ({Oxford University Press}, {New York}, 2006), 6th ed.

\bibitem{HEISS_2001_Eur.Phys.J.D_chirality}
W.~Heiss and H.~Harney, \enquote{The chirality of exceptional points,} Eur.
  Phys. J. D \textbf{17}, 149--151 (2001).

\bibitem{HARNEY_2004_TheEuropeanPhysicalJournalD-AtomicMolecularandOpticalPhysics_Time}
H.~L. Harney and W.~D. Heiss, \enquote{Time {{Reversal}} and {{Exceptional
  Points}},}  \textbf{29}, 429--432 (2004).

\bibitem{BERRY_2006_J.Phys.A:Math.Gen._Proximity}
M.~V. Berry, \enquote{Proximity of degeneracies and chiral points,} J. Phys. A:
  Math. Gen. \textbf{39}, 10013--10018 (2006).

\bibitem{KANE_2005_Phys.Rev.Lett._Quantum}
C.~L. Kane and E.~J. Mele, \enquote{Quantum {{Spin Hall Effect}} in
  {{Graphene}},} Phys. Rev. Lett. \textbf{95}, 226801 (2005).

\bibitem{XIAO_2007_Phys.Rev.Lett._ValleyContrasting}
D.~Xiao, W.~Yao, and Q.~Niu, \enquote{Valley-{{Contrasting Physics}} in
  {{Graphene}}: {{Magnetic Moment}} and {{Topological Transport}},} Phys. Rev.
  Lett. \textbf{99}, 236809 (2007).

\bibitem{HEISS_2008_J.Phys.A:Math.Theor._Chirality}
W.~D. Heiss, \enquote{Chirality of wavefunctions for three coalescing levels,}
  J. Phys. A: Math. Theor. \textbf{41}, 244010 (2008).

\bibitem{RYU_2012_Phys.Rev.A_Analysis}
J.-W. Ryu, S.-Y. Lee, and S.~W. Kim, \enquote{Analysis of multiple exceptional
  points related to three interacting eigenmodes in a non-{{Hermitian
  Hamiltonian}},} Phys. Rev. A \textbf{85}, 042101 (2012).

\bibitem{LEE_2012_Phys.Rev.A_Geometrica}
S.-Y. Lee, J.-W. Ryu, S.~W. Kim, and Y.~Chung, \enquote{Geometric phase around
  multiple exceptional points,} Phys. Rev. A \textbf{85}, 064103 (2012).

\bibitem{HEISS_2015_J.Phys.A:Math.Theor._Resonance}
W.~D. Heiss and G.~Wunner, \enquote{Resonance scattering at third-order
  exceptional points,} J. Phys. A: Math. Theor. \textbf{48}, 345203 (2015).

\bibitem{LIN_2016_Phys.Rev.Lett._Enhanceda}
Z.~Lin, A.~Pick, M.~Lon{\v c}ar, and A.~W. Rodriguez, \enquote{Enhanced
  {{Spontaneous Emission}} at {{Third}}-{{Order Dirac Exceptional Points}} in
  {{Inverse}}-{{Designed Photonic Crystals}},} Phys. Rev. Lett. \textbf{117},
  107402 (2016).

\bibitem{HEISS_2016_J.Phys.A:Math.Theor._model}
W.~D. Heiss and G.~Wunner, \enquote{A model of three coupled wave guides and
  third order exceptional points,} J. Phys. A: Math. Theor. \textbf{49}, 495303
  (2016).

\bibitem{CHEN_2017_Nature_Exceptional}
W.~Chen, {\c S}.~Kaya~{\"O}zdemir, G.~Zhao, J.~Wiersig, and L.~Yang,
  \enquote{Exceptional points enhance sensing in an optical microcavity,}
  Nature \textbf{548}, 192--196 (2017).

\bibitem{HODAEI_2017_Nature_Enhanced}
H.~Hodaei, A.~U. Hassan, S.~Wittek, H.~{Garcia-Gracia}, R.~{El-Ganainy}, D.~N.
  Christodoulides, and M.~Khajavikhan, \enquote{Enhanced sensitivity at
  higher-order exceptional points,} Nature \textbf{548}, 187--191 (2017).

\bibitem{ZHANG_2018_Phys.Rev.X_Dynamically}
X.-L. Zhang, S.~Wang, B.~Hou, and C.~T. Chan, \enquote{Dynamically {{Encircling
  Exceptional Points}}: {{In}} situ {{Control}} of {{Encircling Loops}} and the
  {{Role}} of the {{Starting Point}},} Phys. Rev. X \textbf{8}, 021066 (2018).

\bibitem{PAP_2018_Phys.Rev.A_NonAbelian}
E.~J. Pap, D.~Boer, and H.~Waalkens, \enquote{Non-{{Abelian}} nature of systems
  with multiple exceptional points,} Phys. Rev. A \textbf{98}, 023818 (2018).

\bibitem{DING_2018_Phys.Rev.Lett._Experimentala}
K.~Ding, G.~Ma, Z.~Q. Zhang, and C.~T. Chan, \enquote{Experimental
  {{Demonstration}} of an {{Anisotropic Exceptional Point}},} Phys. Rev. Lett.
  \textbf{121}, 085702 (2018).

\bibitem{ZHANG_2018_Phys.Rev.A_Hybrid}
X.-L. Zhang and C.~T. Chan, \enquote{Hybrid exceptional point and its dynamical
  encircling in a two-state system,} Phys. Rev. A \textbf{98}, 033810 (2018).

\bibitem{ZHONG_2018_Nat.Commun._Winding}
Q.~Zhong, M.~Khajavikhan, D.~N. Christodoulides, and R.~{El-Ganainy},
  \enquote{Winding around non-{{Hermitian}} singularities,} Nat. Commun.
  \textbf{9}, 4808 (2018).

\bibitem{BERRY_2019_J.Opt._Geometry}
M.~V. Berry and P.~Shukla, \enquote{Geometry of {{3D}} monochromatic light:
  Local wavevectors, phases, curl forces, and superoscillations,} J. Opt.
  \textbf{21}, 064002 (2019).

\bibitem{PAPAKOSTAS_Phys.Rev.Lett._optical_2003}
A.~Papakostas, A.~Potts, D.~M. Bagnall, S.~L. Prosvirnin, H.~J. Coles, and
  N.~I. Zheludev, \enquote{Optical manifestations of planar chirality,} Phys.
  Rev. Lett. \textbf{90}, 107404 (2003).

\bibitem{CHEN_2020_Phys.Rev.Research_Scatteringa}
W.~Chen, Q.~Yang, Y.~Chen, and W.~Liu, \enquote{Scattering activities bounded
  by reciprocity and parity conservation,} Phys. Rev. Research \textbf{2},
  013277 (2020).

\bibitem{ZHOU_2016__Tailoring}
H.~Zhou, \enquote{Tailoring light with photonic crystal slabs : From
  directional emission to topological half charges,} Thesis, Massachusetts
  Institute of Technology (2016).

\end{thebibliography}

\end{document}